\documentclass{ws-procs9x6}            

\usepackage[all,knot]{xy}
\xyoption{arc}

\newcommand{\dd}{\mathrm{d}}   
\newcommand{\myxymatrix}[1]{\vcenter{\vbox{\xymatrix{#1}}}}
\newcommand{\CM}{\mathcal{M}}
\newcommand{\CA}{\mathcal{A}}
\newcommand{\CS}{\mathcal{S}}
\newcommand{\CL}{\mathcal{L}}
\newcommand{\CF}{\mathcal{F}}
\newcommand{\dpar}{\partial} 
\newcommand{\der}[1]{\frac{\dpar}{\dpar #1}}   
\newcommand{\lbr}{(\hspace{-0.1cm}(}
\newcommand{\rbr}{)\hspace{-0.1cm})}

\begin{document}
\begin{flushright}
 DIFA 2016
\end{flushright}
\title{Generalized Higher Gauge Theory and M5-brane dynamics.\\
Proceedings: Higher structures in String and M-theory,\\
Tohoku Forum for Creativity, March 7-11 2016}

\author{P. Ritter}

\address{Dipartimento di Fisica e Astronomia, Universit\`a di Bologna,\\
Bologna, Italia\\
$^*$E-mail: pritter@bo.infn.it\\
http://www.fisica-astronomia.unibo.it}

\begin{abstract}
We give a review of truncated L$_\infty$ algebras, as used in the
study of higher gauge theory. These structures are believed to hold
the correct properties to adequately describe gauge theory of extended
objects. We discuss how to construct topological
higher-gauge-invariant theories and how their solutions relate to
multisymplectic geometries. We also show how Courant algebroids fit
into this formalism, so as to be able to study higher gauge
theory on generalized geometric bundles, \textit{i.e.} on $T\Sigma\oplus
T^*\Sigma$, for some space-time $\Sigma$. We will see that via this formalism we can match and explain
a recently proposed M5-brane model, arrived at in a more heuristic way, whose field content seemed
difficult to interpret but finds a natural motivation in this
framework. 
\end{abstract}

\keywords{Strong homotopy algebras, Lie $n$-algebras, higher gauge theory models, generalized
  geometry, Courant algebroids, M5-brane effective dynamics}

\bodymatter
\section{Introduction}
In the spirit of ``Higher structures in String and M-theory'', we will
attempt here a relatively self-contained review of some techniques and
an intriguing explicit example of how higher gauge theory connects
with other novel mathematics introduced for the study of extended
objects.

We know that if we assume the fundamental entities in physics not to be
point-like anymore, we will have to deal with constructing gauge
theories over higher dimensional world-volumes. It has been known for
a long time, however, that it is impossible to make an action (say a
Yang-Mills type model) reparametrization invariant, if we allow for $p$-form
gauge connections, for $p>1$, that take values in a non-abelian symmetry algebra
(for an explicit, physical calculation, see for instance
\cite{Henneaux1986-HENPE}). As was explained in more detail in the lectures
during this workshop, the issue boils down to our inability to
uniquely define ``higher holonomy'' for $(d>1)$-dimensional volumes,
as it is impossible to uniquely assign something like a surface- (or
volume) ordering over which to integrate higher $p$-form
connections. There are obviously many ways to, say, move bits of paths
along a surface, before gluing them back together, that might not
necessarily lead to the same result when integrated over: one needs to
impose that all possibilities, that cover the same world-volume, lead to
the same holonomy assignment. As it turns out,
requiring this sort of equivalence, corresponds precisely to the set
of defining rules of higher categories. This may not be surprising, as
we are really just restating, in different languages, our need for a
certain amount of associativity between different ``products'' in a
very general set-up. 

The power of category theory comes in at the next
step: since we can use functors $F$ to map between categories $\mathcal{C}_1\stackrel{F}{\rightarrow}\mathcal{C}_2$, and these
have to preserve the properties of all the maps of both
$\mathcal{C}_1$ and $\mathcal{C}_2$, the various components of $F$
will in turn have to satisfy a very specific set of rules. For our
intents and purposes, the categories to map between are usually a
space-time manifold on the one side, and an interior symmetry
group-like structure
on the other. The functor $F$ between them will contain the necessary information
about the gauge connections' properties, according to how high the two categories
are. In the familiar example of ordinary gauge theory, we would be
mapping between the \textit{path 1-groupoid} associated to a manifold
(where the objects are points and morphisms are paths, or world-lines, between them,
roughly speaking) and a gauge group $G$, which in turn can be seen as
a \textit{1-groupoid}, with just the one object $G$ and the group
elements as morphisms. One then has a functor that associates a group element to each
closed path $\gamma$ on the manifold, via the familiar holonomy map:
\[
F:\quad \gamma\longrightarrow \mathcal{P}\left(\exp{\int_\gamma
    A}\right)\quad\in G~,
\]
where $\mathcal{P}$ stands for the path-ordering and $A$ is a 1-form valued in the Lie algebra of $G$. Requiring
that $F$ be a functor (that is, preserves the associative morphisms of
both underlying categories) in fact imposes on $A$ all the properties
that define a connection (see \cite{Baez:2010ya} for a pedagogical introduction).

Moving up one step, world-surfaces would be described, roughly
speaking, by the \textit{2-category} consisting of points, paths and
surfaces between them, while its gauge symmetry structure would be
expected to be a \textit{Lie 2-group}. These higher categories are
endowed with \textit{2-morphisms}, as well as morphisms, whose various
mixed composition rules also have to be preserved by what will here be
a \textit{2-functor}. A
consistent analogue of the holonomy map now has to be constructed out of
a pair of gauge fields $(A_\mu,\,B_{\mu\nu})$, a 1- and a
2-form respectively, each valued in one of the two components of the
\textit{Lie 2-algebra} corresponding to the Lie 2-group in
question. These structures, in turn, have to satisfy their own
specific set of rules, that we will elucidate in the first section
below. Furthermore, $A$ and $B$ can be shown to have to satisfy the so-called \textit{vanishing fake
  curvature condition}, \textit{i.e.} that $F-t(B)=0$, where $t$ is a
map between the two components of the 2-algebra. For details on this
see C. S\"amann's lectures, part of these same Proceedings, or \cite{math/0511710,Baez:2010ya}.

In what follows we will not delve into the details of this
motivation. We will however present the definitions and salient
properties of the mathematical tools that are used for higher gauge
theory. We will explain how one can describe higher Lie algebras, as well
as some examples of them that have already been encountered in
physics. We will also show how one can construct general topological
models based on higher symmetry structures and how these relate to a
higher analogue of the Poisson algebra on symplectic manifolds,
\textit{i.e.} to the Lie $n$-algebras on $n$-plectic spaces. 

In section \ref{relation to courant} we will show explicitly
how Courant algebroids, as they appear in generalized complex geometry
and in double field theory, come equipped with their own Lie
2-algebras. In fact, we will see that they contain the 2-algebra
structure of a 2-plectic manifold. We then consider the space-time side of
our connection functor to be already generalized, to $T\Sigma\oplus
T^*\Sigma$, while allowing for a general 2-algebra on the internal
symmetry side. Finally, in section \ref{M5-brane model} we will show how the flatness conditions on the higher
connections in this setup precisely reproduce the equations of motion
proposed recently, in
\cite{Lambert:2010wm}, for the effective dynamics of M5-branes. There, a
different generalization of Lie algebras is used, but we will again
see that it is just another example of a special, \textit{strict}, 2-algebra.

\section{Mathematical tools}
Let us first introduce the arsenal of mathematical tools that will be
needed in the following sections. Here we will give two equivalent
definitions of truncated strong homotopy algebras (denoted L$_\infty$-algebras), or Lie $n$-algebras,
each of which we will see to be useful in different contexts. 

\subsection{Lie $n$-algebras}
\begin{definition}
An {\em $L_\infty$-algebra} or {\em strong homotopy Lie algebra} is a graded vector space $L=\oplus_i L_i$ endowed with $n$-ary multilinear totally antisymmetric products $\mu_n$, $n\in\mathbb{N}^*$, of degree $(2-n)$, that satisfy homotopy Jacobi identities, cf.\ \cite{Lada:1992wc,Lada:1994mn,0821843621}. These identities read as
\begin{equation*}\label{eq:homotopyJacobi}
 \sum_{i+j=n}\sum_\sigma\chi(\sigma;l)(-1)^{i\cdot j}\mu_{j+1}(\mu_i(l_{\sigma(1)},\cdots,l_{\sigma(i)}),l_{\sigma(i+1)},\cdots,l_{\sigma(i+j)})=0
\end{equation*}
for all $n\in \mathbb{N}^*$, where the sum over $\sigma$ is taken over
all $(i,j)$ unshuffles. A permutation $\sigma$ of $i+j$ elements is
called an {\em $(i,j)$-unshuffle}, if the first $i$ and the last $j$
images of $\sigma$ are ordered: $\sigma(1)<\cdots<\sigma(i)$ and
$\sigma(i+1)<\cdots<\sigma(i+j)$. Moreover, the {\em graded Koszul
  sign} $\chi(\sigma;l)$, for $l=(l_1,\ldots,l_n)$ and $l_i\in L$ is defined via the equation
\begin{equation*}
 l_1\wedge \cdots \wedge l_n=\chi(\sigma;l)\,l_{\sigma(1)}\wedge \cdots \wedge l_{\sigma(n)}
\end{equation*}
in the free graded algebra $\wedge (l_1,\cdots,l_n)$, where $\wedge$
is considered graded antisymmetric.
\end{definition}

\textit{Truncated} strong homotopy Lie algebras are
concentrated in degrees $(-n+1),\ldots,0$, so that\ $L_i=*$
for $i\notin [-n+1,\ldots,0]$. Consequently, because of their grading,
the $\mu_k$ products will vanish for $k>(n+1)$. These truncated
L$_\infty$ algebras are believed to be categorically equivalent to semi-strict Lie
$n$-algebras\footnote{This has as yet only been proven for $n=2$, nonetheless we will
  continue to use both terms interchangeably for the remainder of this
  paper.}, and are therefore expected to be the correct
infinitesimal symmetry structure for gauge theories of extended
objects.

Specifically, we will be interested in the case of semi-strict Lie
$2$-algebras, which will be given by the 2-term real vector-space complex 
\[
L:\ \ V\stackrel{\mu_1}{\longrightarrow}W\stackrel{\mu_1}{\longrightarrow}0~,
\]
where here $L_{-1}\equiv V$ and $L_0\equiv W$. 

The $n=2$ example is of interest when studying, for instance, gauge
theory over the world-surface of a 1-dimensional object, such as a
string. In particular, when searching for an effective gauge theory as
seen by an M2-brane intersecting a stack of extended objects, for
which we expect a non-abelian internal symmetry group \cite{Ritter:2013wpa}.

The homotopy product
$\mu_1$ has degree 1 and squares to zero, while the grading also
imposes that
\begin{equation*}
\begin{aligned}
\mu_1(w)&=0~,~~~
 \mu_2(v_1,v_2)=0~,\\
\mu_3(v_1,v_2,v_3)&=\mu_3(v_1,v_2,w)=\mu_3(v_1,w_1,w_2)=0~.
\end{aligned}
\end{equation*}
The 2-algebra's non-vanishing higher products
satisfy the following higher Jacobi identities: 
\begin{equation*}
\begin{aligned}
 \mu_1(\mu_2(w,v))&=\mu_2(w,\mu_1(v))~,~~~\mu_2(\mu_1(v_1),v_2)=\mu_2(v_1,\mu_1(v_2))~,\\
 \mu_1(\mu_3(w_1,w_2,w_3))&=-\mu_2(\mu_2(w_1,w_2),w_3)-\text{cyclic}(w_1,w_2,w_3)~,\\
 \mu_3(\mu_1(v),w_1,w_2)&=-\mu_2(\mu_2(w_1,w_2),v)-\text{cyclic}(w_1,w_2,v)~,
\end{aligned}
\end{equation*}
where $v_i\in V$ and $w_i\in W$ have degrees -1 and 0 respectively.

The equalities above show how the elements in $V$ and $W$ mix in a non-trivial
way: indeed, only when $\mu_3=0$, we are just describing a \textit{differential
crossed module} of actual Lie algebras, given by $(W,\, V,\, \mu_1,\alpha)$, where the action
$\alpha$ of $W$ on $V$ is given by the product $\mu_2(w,v)$. It is clear from the last two lines that, for non-vanishing $\mu_3$,
the Jacobi identity of traditional Lie algebras is violated in a
controlled way, by a $\mu_1$-exact term. 

We have one further identity coming from definition
(\ref{eq:homotopyJacobi}):
\begin{equation*}
\begin{aligned}
 \mu_2(\mu_3(w_1,&w_2,w_3),w_4)-\mu_2(\mu_3(w_4,w_1,w_2),w_3)+\mu_2(\mu_3(w_3,w_4,w_1),w_2)\\
 & -\mu_2(\mu_3(w_2,w_3,w_4),w_1)=\\
 \mu_3(\mu_2(w_1,&w_2),w_3,w_4)-\mu_3(\mu_2(w_2,w_3),w_4,w_1)+\mu_3(\mu_2(w_3,w_4),w_1,w_2)\\
 -\mu_3(\mu_2(w_4&,w_1),w_2,w_3)
 -\mu_3(\mu_2(w_1,w_3),w_2,w_4)-\mu_3(\mu_2(w_2,w_4),w_1,w_3)~,
\end{aligned}
\end{equation*}
specifying how the ternary product $\mu_3$ mixes with $\mu_2$.

Just like with Lie algebras and the Chevalley-Eilenberg complex, here
too we have an equivalent dual description of the structure, via
$NQ$-manifolds (as introduced in \cite{Severa:2001aa}):
\begin{definition}
An $NQ$-manifold is a $\mathbb{N}$-graded manifold 
\[
\mathcal{M}=M_0\leftarrow M_1\leftarrow M_2\leftarrow\cdots~,
\]
endowed with a degree 1,
nilpotent, differential operator $Q$:
\begin{equation}
  Q=\sum_i \frac{1}{i!}m^B_{C_1\cdots C_i}Z^{C_1}\cdots
  Z^{C_i}\frac{\partial}{\partial Z^B}~,
 \label{eq:3}
\end{equation}
where $\{Z^{C_k}\}$ are coordinates of degree $[Z^{C_k}]$ parametrizing $\mathcal{M}$, and $\sum_{l=1}^i[Z^{C_l}]=[Z^B]+1$.
\end{definition}
Note that in the above definition the components $M_i$ have
positive degree, while in the previous L$_\infty$-algebra definition we
started with a complex of negatively graded vector spaces. This is a
matter of convention: one could easily redefine the $L_i$ components
to have positive grading, and $\mu_1$ to map ``downward'', but we prefer to stick to these choices so
as to stay in line with what is most commonly used in the literature.

Requiring the nilpotency of the general operator $Q$ in (\ref{eq:3})  yields a set of
conditions on the ``structure coefficients'' $m^B_{C_1\cdots C_i}$, which
are just the dual equivalent of the higher Jacobi identities we obtain
from eqn. (\ref{eq:homotopyJacobi}). Indeed, if we take a manifold
$\mathcal{M}$ which has no degree zero component, $M_0=*$, and
make the identification
\[
\mu_k(\tau_{C_1},\ldots, \tau_{C_k})=m^B_{C_1\cdots
    C_k}\tau_B~,
\]
where $\tau_A$ is a basis for $\mathcal{M}$, and we assign degrees
$[\tau_A]=1-[Z^A]$ to adjust for the inversion of the grading between
the L$_\infty$ complex and the $NQ$-manifold definitions mentioned above.
Requiring $Q^2=0$
will translate to the correct higher homotopy structure for the
$\mu_k$. When $M_0\neq*$, the construction of the homotopy
products is more subtle, requiring the use of derived coalgebra
techniques. In this case one is describing \textit{Lie $n$-algebroids},
which are just the categorification of traditional Lie
algebroids. \\
Let us consider again the 2-algebra example: take $\mathcal{M}=W[1]\oplus V[2]$, with coordinates $\{ w^a,\,v^i\}$
  of degrees $(1,\,2)$. A general degree-1 differential operator $Q$ is given by
  \begin{equation}
    Q=\left(-\frac12 f^a_{bc}w^bw^c
      -t^a_iv^i\right)\frac{\partial}{\partial
      w^a}+\left(\frac16
      h_{abc}^iw^aw^bw^c-g_{aj}^iw^av^j\right)\frac{\partial}{\partial
      v^i}~.
\label{Q for n-algebra}
  \end{equation}
Making the following identifications:
\begin{equation*}\begin{array}{c}
  \mu_1(\lambda_i)=t^a_i\tau_a~,\quad
  \mu_2(\tau_a,\tau_b)=f^c_{ab}\tau_c~,\quad
  \mu_2(\tau_a,\lambda_i)=g_{ai}^j\lambda_j~,\\ \ \\
\mu_3(\tau_a,\tau_b,\tau_c)=h_{abc}^i\lambda_i~,
\end{array}
\end{equation*}
for $\{\tau_a\}$ spanning $W$ and $\{\lambda_i\}$ spanning $V$, one
can easily check that requiring $Q^2=0$ yields the correct higher
Jacobi identities.

\subsection{Multisymplectic spaces}\label{multisymplectic spaces}
Let us now look at a specific example of a realisation of these higher
homotopy structures. To do this, we look at a higher analogue of the
symplectic structure on even-dimensional manifolds and the Lie algebra
product it induces. 
\begin{definition}
A \textit{multisymplectic manifold}, or \textit{$n$-plectic manifold},
is a manifold $M$ endowed with an $(n+1)$-form $\omega$ that is
\begin{itemize}
\item closed, \textit{i.e.} $\dd\omega=0$;
\item non-degenerate, \textit{i.e.}
  $\iota_X\omega=0\,\Leftrightarrow\, X=0$, where $X$ is a vector
  field on $M$.
\end{itemize}
\end{definition}
Just like in symplectic geometry, we can use this structure to define
\textit{Hamiltonian vector fields} $X_\alpha$ corresponding to
$(n-1)$-forms $\alpha$ via
\[
\dd\alpha=-\iota_{X_\alpha}\omega~.
\]
This immediately suggests how to define higher order products
(cf. \cite{Barnich:1997ij, Rogers:2010nw, Rogers:2011zc}):
\begin{definition}
The \textit{strong homotopy algebra of local observables} (or
\textit{shlalo} for short) of $(M,\,\omega)$, denoted by $\Pi_n$,
is given by the vector-space complex
 \begin{equation*} 
    L:\quad C^\infty(M)\stackrel{\pi_1}{\longrightarrow}\Omega^1(M) \stackrel{\pi_1}{\longrightarrow}\cdots \stackrel{\pi_1}{\longrightarrow}\Omega^{n-1}(M)~,
  \end{equation*}
together with the brackets for $f\in
  C^\infty(M)~,\, \alpha_l\in\Omega^{n-1}(M)$:
\begin{equation*}
  \pi_1(f)=\dd f~, \quad \pi_k(\alpha_1,\ldots,\alpha_k)=(-1)^{\binom{k+1}{2}}
  \iota_{X_{\alpha_1}}\cdots \iota_{X_{\alpha_k}}\omega~.
\end{equation*}
\end{definition}
$\Pi_n=(M,\omega, \pi_k)$ is clearly a Lie $n$-algebra. It is not
exactly the higher analogue of a Poisson algebra, because there is no
obvious way to define an associative product between observables
(between two 1-forms, for example) that respects the product
structure. Nonetheless, from a physics point of view, it is of course
tempting to expect these objects to be the 'classical limit' of the
$n$-algebra structure of some quantum theory. Alternatively, one might
expect $\Pi_n$ to be the starting point to quantizing an
$(n+1)$-dimensional world-volume. Indeed, one can relate these
$n$-algebras to Nambu-Poisson structures of rank $(n+1)$ (see
\cite{Ritter:2015ymv}). For completeness' sake, let us note that Nambu-Poisson
structures themselves are expected to quantize to Lie $n$-algebras,
albeit in a quite complicated way (see \cite{DeBellis:2010pf} and the
references therein). One type of structure
that was hoped to encode the quantum behaviour of extended objects was the triple bracket introduced by Bagger, Lambert and
Gustavsson to describe stacks of M2-branes \cite{Bagger:2006sk,
  Gustavsson:2007vu}. These
\textit{BLG 3-Lie-algebras} are in fact \textit{strict} Lie 2-algebras (\textit{i.e.} they have
$\mu_3=0$, so they are differential crossed modules of pairs of actual
Lie algebras), so the strong homotopy algebra language may be
the correct approach for the necessary generalizations. Interestingly, the
ABJM model for M2-branes\cite{Aharony:2008ug} can also be shown to be a higher gauge theory
\cite{Palmer:2013ena}.

We will show an explicit
example of a BLG 3-Lie algebra and its corresponding strict 2-algebra
in the last section,
but for a more general discussion of the correspondence we recommend
\cite{Palmer:2012ya}.

\subsection{Symplectic $NQ$-manifolds}
We are still missing a fundamental ingredient for the construction of
physical actions: that is an invariant metric via which to pair Lie $n$-algebra
valued fields. Such an inner product is best introduced in the
$NQ$-manifold framework. 
\begin{definition}
A \textit{symplectic $NQ$-manifold} is an $NQ$-manifold $(\mathcal{M},\,Q)$
endowed with a closed, non-degenerate, 2-form
    $\varpi$, of ``ghost'' degree $p=n+1$, invariant under $Q$:
    \begin{equation*}
      \varpi=\tfrac12 \varpi_{AB}\dd Z^A\wedge\dd Z^B~,
      \quad\text{s.t.}\quad \mathcal{L}_{Q}\varpi=0~.
    \end{equation*}
Again, we have denoted with $Z^A$ the coordinates on $\mathcal{M}$ and
$\mathcal{L}$ indicates the Lie derivative. As usual, $n$ here is the
degree of the highest weight coordinate on $\mathcal{M}$.
\end{definition}
Since $\varpi$ is non-degenerate, its inverse can be used to induce a
bilinear graded symmetric inner product: $\{-,-\}_\varpi:\,C^\infty(\mathcal{M})\times
C^\infty(\mathcal{M})\rightarrow C^\infty(\mathcal{M})$. As in
symplectic geometry, each function $F$ on $\mathcal{M}$ has a
corresponding vector defined by $\dd F=-\iota_{V_F}\varpi$, and one
sets
\[
\{F,\,G\}_\varpi:=\iota_{V_F}\iota_{V_G}\varpi~.
\]
This structure also allows us to find the ``Hamiltonian''
$\mathcal{S}$ associated
to the nilpotent $Q$ operator, since $Q(F)=\{\mathcal{S},\,F\}_\varpi$, which
squares to zero in the bracket: $\{\mathcal{S},\mathcal{S}\}_\varpi=0$.

On the dual side, for the Lie $n$-algebra $L$ defined by the
symplectic $NQ$-manifold, $\varpi$ translates to a
metric on the
vector space, $(-,-):\,L\times L\rightarrow \mathbb{R}$. In
particular, for $l_k\in L$, it will have the following symmetry and
invariance under the $k$-ary products:
\begin{equation*}\begin{array}{l}
  (l_1,l_2)=(-1)^{|l_1|+|l_2|}(l_2,l_1)~,\\[1.5ex]
\left(\mu_k(l_1,\ldots,l_k),l_0\right)=(-1)^{k+|l_0|(|l_1|+\cdots+|l_k|)}\left(\mu_k(l_0,\ldots,l_{k-1}),l_k\right)~.
\end{array}
\end{equation*}
This is usually referred to as a \textit{cyclic metric} (for more
details and the original reference see \cite{Kontsevich:1992aa,
  Igusa:2003yg, Zwiebach:1992ie, 0821843621}), while
$(L,\,(-,-))$ is now a \textit{metric Lie $n$-algebra}. 

We now have all the elements in our mathematical tool-box to start
constructing gauge-theory actions. We will be dealing with multiplets
of gauge connections, valued in truncated L$_\infty$-algebras $L$ endowed
with a cyclic metric induced by $\varpi$. The latter, being compatible
with the $Q$-structure on the space dual to $L$, allows us to easily
select only gauge-invariant objects, but also to then apply
variational principles to our model.

\section{Topological $n$-algebra models}
\subsection{Construction}\label{Construction}
We often refer to the following as a generalization of the AKSZ method
for constructing actions\cite{Alexandrov:1995kv}, but it is in fact inspired by the
work of Atiyah \cite{Atiyah:1957} and later applications by various
authors \cite{Bojowald:2004wu, Kotov:2007nr, Gruetzmann:2014ica,
  Sati:2008eg, Fiorenza:2011jr}. We start with the following diagram:
  \begin{equation*}
  \xymatrixcolsep{5pc}
  \myxymatrix{
  & T[1]\CM \ar@{->}[d]^{\pi} \\
  T[1]\Sigma \ar@{->}[ur]^{f} \ar@{->}[r]^{a} & \CM
  }
\end{equation*}
of an $NQ$-manifold $\mathcal{M}$, representing the internal symmetry
algebra of our theory, its tangent bundle $T[1]\mathcal{M}$ and the
tangent bundle of space-time $T[1]\Sigma$. The number $[k]$ in square
brackets indicates a shift in the degree of the coordinates of that
space. Each of these spaces comes with a $Q$-structure:
$Q_\Sigma=\dd_\Sigma$ on the space-time and $Q_{\mathcal{M}}$ will be the
usual dual, or higher Chevalley-Eilenberg, operator on a Lie
$n$-algebra (see definition (\ref{Q for n-algebra})). On $T[1]\mathcal{M}$ there
are two degree-1 differential maps, whose sum gives the full operator:
$Q_{T\mathcal{M}}=\bar\dd+\mathcal{L}_{Q_{\mathcal{M}}}$, where
$\bar\dd$ is just a degree-shift operator.  \\
The degree-preserving map $a:\,T[1]\Sigma\rightarrow \CM$ will be referred to as a
connection. The map $f$ has been introduced because, as opposed to
$a$, it does commute with the $Q$-structures and is therefore a
$Q$-morphism: 
\[
f^*(\pi^*h)=a^*(h)~,
\quad\text{and}\quad f^*(\bar\dd h)=\left(\dd_\Sigma\circ a^*-a^*\circ Q_\CM\right)(h)~,
\]
where $h\in C^\infty(\mathcal{M})$. Clearly we have $f^*\circ
Q_{T\mathcal{M}}=Q_\Sigma\circ f^*$. Explicitly, for the coordinate
$Z^K$ on $\CM$, let us call 
\[
\mathcal{A}^K=\tfrac{1}{K!}\mathcal{A}^K_{\mu_1\cdots\mu_K}\dd
x^{\mu_1}\wedge\cdots\wedge\dd x^{\mu_K}:=a^*(Z^K)~,
\]
where we have called the weight $[Z^K]\equiv K$, so that $Z^K$ pulls back to a $K$-form.
The
pullback along $f$ of $\bar\dd Z^K$, or, equivalently, the failure of
$a$ to be a $Q$-morphism, then gives the \textit{higher (fake) curvature} of $\mathcal{A}^K$ :
\begin{equation}
\mathcal{F}^K=\tfrac{1}{(K+1)!}\mathcal{F}^K_{\mu_1\cdots\mu_{K+1}}\dd x^{\mu_1}\wedge\cdots\wedge\dd
x^{\mu_{K+1}}:=\left(\dd_\Sigma\circ a^*-a^*\circ Q_{\CM}\right)(Z^K)~,
\label{curvature def}
\end{equation}
which has form-degree one more than the weight of the coordinate $Z^K$. 

Going back to our 2-algebra example, with $\CM=W[1]\oplus V[2]$,
parametrized by $\{w^a\},\,\{v^i\}$, graded 1 and 2 respectively, we
have
\[
a^*(w^a)=A^a=A^a_\mu\dd x^\mu~,\qquad a^*(v^i)=B^i=\tfrac12B^i_{\mu\nu}\dd
x^\mu\wedge\dd x^\nu~,
\] 
so that $\mathcal{A}=(A,\,B)$ form a \textit{2-connection}, with
(fake) curvatures
\begin{equation*}\left\{\begin{array}{l}
\mathcal{F}^a=\dd_\Sigma
A^a+\tfrac12\mu_2(A,A)^a-\mu_1(B)^a\\[1.5mm]
 \mathcal{F}^i=\dd_\Sigma B^i+\mu_2(A,B)^i-\tfrac16\mu_3(A,A,A)^i~~.
\end{array}\right.
\end{equation*}
We would now like to write down actions with fields valued in a Lie
$n$-algebra, that are invariant under the internal symmetry. This
ultimately means that we are looking to pull back $Q$-invariant
polynomials on $T[1]\CM$ to our space-time manifold $T[1]\Sigma$. We
further want to respect the cohomology from the $n$-algebra side, that
is we want $Q_{T\CM}$-exact terms to pull back to $\dd_\Sigma$-exact
objects. The most obvious exact invariant polynomial on $T[1]\CM$ is of
course its symplectic structure $\varpi$. On $T[1]\CM$ it is given by
$\varpi=\varpi_{AB}Q_{T\CM}Z^AQ_{T\CM}Z^B=\varpi_{AB}(\bar\dd
+Q_\CM)Z^A(\bar\dd +Q_\CM)Z^B$. Let us consider its
``non-covariant'' version, $\hat\varpi=\hat\varpi_{AB}\bar\dd
Z^A\bar\dd Z^B$, which is also $Q_{T\CM}$-invariant (since
$\iota_Q\varpi=\iota_Q\hat\varpi=-\dd\mathcal{S}$, is exact). Because of the definition of higher curvature from
above, it is clear that this object should pull back to
the product of $\mathcal{F}^a$ and $\mathcal{F}^i$. So let us look for
the ``potential'' giving rise to this polynomial: some $\chi$, such
that $\hat\varpi=(\bar\dd +\CL_{Q_\CM})\chi$. We again want
$Q_\CM$-invariance, which translates to requiring that if $\chi$
restricted to $\CM$ (projected via $\pi$) is given by some function
$\kappa\in C^\infty(\CM)$, then $Q_\CM(\kappa)=0$. It can be easily
verified that this potential $\chi$ is given by
\[
\chi=\varpi_{AB}Z^A\bar\dd Z^B-\mathcal{S}~,
\]
which is referred to as the \textit{Chern-Simons element}, for
topological field theories. Under the pullback $f^*$, this element
will give us an action invariant under the gauge $n$-algebra:
\[
S=\int_\Sigma f^*{\chi}=\int_\Sigma\left[ (\CA,
    \CF)+a^*(\CS)\right],
\]
where we recall that $(-,-)$ is the cyclic metric on the algebra induced by
$\varpi$, and $\mathcal{S}$ is the ``Hamiltonian'' to
$Q_\CM$. Regrouping all the higher gauge connections into a single
field
\[
\phi=\sum_A(\pm
  a^*(Z^A))~,
\]
where the signs can be freely chosen or reabsorbed into the fields,
the equations of motion will be given by
\[
\dd_\Sigma\phi+ \sum_k\tfrac{(-1)^{\sigma_k}}{k!}\mu_k(\phi,\ldots,\phi)=0~,
\]
where the signs $\sigma_k$ will depend on the choices made for $\phi$.
These can also be written as just $\mathcal{F}^A=0$, for each degree
of the components of $\CM$, \textit{i.e.} they are just \textit{higher
flatness conditions}, also referred to as the
\textit{higher Maurer-Cartan equations} (for more details on this
set-up, see\cite{Ritter:2015ymv}). As it happens, these are also
the conditions that category theoretical considerations require so as
to have a well-defined concept of higher holonomy on $n$-dimensional
world-volumes, as mentioned in the introduction. They are therefore considered to be of fundamental
importance for any consistent higher gauge theory.

\subsection{Solutions}\label{n-plectic solutions}
So far, everything has been very abstract, as we have been dealing in
formal products $\mu_k$ and general $n$-algebras. We have reached the
higher flatness conditions, but we do not as yet have any more
intuitive picture of what is going on. This is where the
$n$-plectic spaces we discussed in section \ref{multisymplectic
  spaces} will come in useful. First, however, let us see how to go
about solving our higher Chern-Simons models. 

From the study of the IKKT model (see \cite{Ishibashi:1996xs, Aoki:1998bq}), we know that the
0-dimensional reduction of 10-dimensional SYM theory, as a matrix
model, looks like a na\"ive quantization of type IIB string theory,
when written in a particular gauge. The embedding coordinates
$X^\mu$ quantize to the matrix valued fields $A^\mu$, that are to
satisfy variational equations of the type
$[A_\mu,[A^\mu,A^\nu]]=0$. Such conditions are clearly solved by the
Moyal plane $\mathbb{R}^{2k}_{\theta}$:
\[
[A^\mu,\,A^\nu]=:[\hat X^\mu,\hat X^\nu]\sim\theta^{\mu\nu}~,
\]
for a constant $\theta$, that is, by quantized embedding coordinates that satisfy the
Heisenberg algebra. This kind of solution is to be interpreted as a
quantum space-time emerging out of the non-perturbative model,
carrying with it the information about how non-commutative the
geometry is that the string sees at a high enough energy limit. We
follow exactly the same philosophy for our higher gauge theories: we
consider our higher CS-theory as if it were the analogue of
SYM, as the effective theory for stacks of branes\footnote{Of course we are talking about a topological
  theory versus SYM, but we are only after a very simple
  toy-model analysis, to highlight some interesting features of the
  models, not an actual higher analogue of YM-theory.}; we reduce our model to 0 dimensions, expecting it to be
the high-energy limit, non-perturbative version of some classical
theory for extended objects; we thus check if the $\Pi_n$ version
of a higher Poisson algebra of an $(n+1)$-dimensional object respects
the reduced equations of motion of our theory. 

From the previous subsection, it is easy to read off the equations of
motion for a Lie 2-algebra model, reduced to zero dimensions:
 \begin{equation}
    \begin{array}{l}
      \CF^0_{ij}=\tfrac12 \mu_2(A_i,A_j)-\mu_1(B_{ij}) \stackrel{!}{=}0\\
     \CF^0_{ijk}=\tfrac16
      \mu_3(A_i,A_j,A_k)+\mu_2(A_i,B_{jk})\stackrel{!}{=}\epsilon_{ijk}~,
    \end{array}
\label{2-algebra 0-dim eom}
 \end{equation}
where we allow the 3-form curvature not to vanish, because
this condition is not actually needed for well-defined holonomy on a
2-dimensional surface. Just like in non-commutative Yang-Mills theory, this \textit{twist} of the homotopy
Maurer-Cartan equations allows for interesting non-commutative
solutions, as we will see now. 

Consider the shlalo of $\mathbb{R}^3_\omega$: on
3-dimensional space, a 2-plectic form is obviously given by the
volume form
\[
\omega=\dd\text{vol}=\tfrac16\epsilon_{ijk}\dd x^i\wedge\dd
  x^j\wedge\dd x^k~,
\]
while the shlalo products are given by
\[
\pi(f)=\dd f~, \qquad
  \pi_2(\alpha,\beta)=\iota_{X_\alpha}\iota_{X_\beta}\omega~,\qquad \pi_3(\alpha,\beta,\gamma)=\iota_{X_\alpha}\iota_{X_\beta}\iota_{X_\gamma}\omega~.
\]
If we choose as a basis of Hamiltonian vector fields
$X_{A_i}=\tfrac{\partial}{\partial x^i}$, the corresponding
Hamiltonian 1-forms are given by 
\[
 A_i=\tfrac12\epsilon_{ijk}x^j\dd x^k~.
\]
We also need a basis for the functions, which here come from the
pullback of degree 2 objects, so they can be given by the 2-forms
$B_{ij}=\epsilon_{ijk}x^k$. With these choices, it is easy to verify
that $(\mathbb{R}^3_\omega,\,\omega,\,\Pi_2)$ solves the equations of motion
(\ref{2-algebra 0-dim eom}).

This example can be easily generalized to higher $n$, higher dimensional
spaces but also different non-commutative geometries, if we
allow the action to contain other gauge-invariant ``deformation
terms'' (some examples for more heuristically constructed actions can
be found in \cite{Ritter:2013wpa}). Furthermore, the
0-dimensional actions can be expanded around the
solutions and give back what will look like BF-theory on the
non-commutative background (again, see\cite{Ritter:2013wpa} for a
detailed example).

\section{Relation to Courant algebroids}\label{relation to courant}
Earlier we saw how $n$-plectic manifolds can carry a Lie $n$-algebra,
as an explicit example of how these structures might appear
in the physics of extended objects. It may not be too surprising to
find that another ``generalized'' structure, introduced for the study of the novel symmetries seen by 1-dimensional
strings, that is T-duality, is also just another example of a
higher algebraic structure. Indeed, we will see in what follows how the Courant
algebroid, one of the salient features of generalized
complex geometry\cite{Gualtieri:2003dx}
and of double field theory\cite{Hull:2009mi}, can be seen as the Lie 2-algebra carried
by a particular Lie 2-algebroid. 

\subsection{Courant algebroids as $NQ$-manifolds}
It was first shown in \cite{Roytenberg:1998vn} that the Courant algebroid structure is
an example of a Lie 2-algebra. In the following we give a derivation
of this fact via the $NQ$-manifold language (as was done in \cite{Roytenberg:2002nu}), which is powerful because
it is easily generalizable to higher dimensional differential forms
(rather than just the 1-forms from $TM\oplus T^*M$), but also because
it can be used to construct actions via $NQ$-manifold morphisms, as we
will see in section \ref{generalized
  higher gth}.

Consider the symplectic $NQ$-manifold $\CM=T^*[2]T[1]\Sigma$, for some
degree 0 manifold $\Sigma$. We denote the coordinates on $\CM$ by
$(x^\mu,\,\xi^\mu,\,\xi_\mu,\,p_\mu)$ so that their weights are $(0,
\,1,\,1,\,2)$ respectively, making this a Lie \textit{2-algebroid}, with
non-vanishing \textit{body} $\Sigma$ of degree 0. The non-degenerate symplectic structure and
the nilpotent $Q$ will be given by
\begin{equation*}
\begin{aligned}
\varpi&=\dd x^\mu\wedge\dd
p_\mu+\dd\xi^\mu\wedge\dd\xi_\mu~,\\
Q&=\xi^\mu\frac{\partial}{\partial
    x^\mu}+p_\mu\frac{\partial}{\partial
    \xi_\mu}+\frac12 H_{\mu\nu\rho}\xi^\mu\xi^\nu\frac{\partial}{\partial
      \xi_\rho}+\frac{1}{3!}\partial_\mu
      H_{\nu\kappa\lambda}\xi^\nu\xi^\kappa\xi^\lambda\frac{\partial}{\partial
        p_\mu}~,
\end{aligned}
\end{equation*}
where $H$ is a closed 3-form introduced for generality. One could even
introduce more structure, \textit{e.g.} a 3-vector $\tilde
H^{\mu\nu\rho}$, or various mixed tensors, but these all go beyond the
scope of our present analysis. 

Consider now the functions over the degree-1 component of $\CM$:
\[
e:=X^\mu\xi_\mu+\alpha_\mu\xi^\mu\quad\in C^\infty(M_1)~,
\]
and define a metric on this space via the Poisson bracket induced by
$\varpi$:
\[
(e_1,e_2):=\tfrac12\{e_1,\,e_2\}_\varpi=\tfrac12\left(X^\mu\beta_{\mu}+Y^\mu\alpha_{\mu}\right)~,
\]
for $e_1=X+\alpha$ and $e_2=Y+\beta$. At this point, it is worth using
the identifications $\xi^\mu\sim\dd x^\mu$ and
$\xi_\mu\sim\partial_\mu$, to make explicit how $e=X+\alpha\in
T\Sigma\oplus T^*\Sigma$. The above metric therefore describes the usual pairing
from generalized complex geometry:
\[
(X+\alpha,\,Y+\beta)=\tfrac12\left(\iota_X\beta+\iota_Y\alpha\right)~.
\]
It so happens that we can also introduce here an antisymmetric product
on $M_1$, constructed with the Hamiltonian function $\Theta$
corresponding to $Q$ itself\footnote{As usual, we can define
  $Q\,F=\{\Theta,\,F\}_\varpi$, so that $\{\Theta,\Theta\}_\varpi=0$.}:
\[
\mu_2(e_1,e_2)=\tfrac12(\{\{\Theta,e_1\},e_2\}_\varpi-\{\{\Theta,e_2\},e_1\}_\varpi)~,
\]
which can be easily verified to translate to
\[
 \mu_2(X+\alpha,Y+\beta)=[X,Y]+\CL_X\beta-\CL_Y\alpha-\tfrac12\dd(\iota_X\beta-\iota_Y\alpha)
   +\iota_X\iota_Y H~,
\]
that is the antisymmetric version of the twisted Courant bracket for the
Courant algebroid $T\Sigma\oplus T^*\Sigma$. We have kept antisymmetry here, at
the cost of the Jacobi identity, since here
\[
\mu_2(e_1,\mu_2(e_2,e_3))+\text{cycl.}=\tfrac13\left((e_1,\mu_2(e_2,e_3))+\text{cycl.}\right)=:\dd\left(\mu_3(e_1,e_2,e_3)\right)~.
\]
That is, the associativity of the algebra product is violated by an
exact term, the argument of which we call $\mu_3$. In addition, we can
call $\mu_1$ the action of $Q$ on $f(x)\in C^\infty(M)$:
\[
\mu_1\left(f(x)\right):=\{\Theta,\,f(x)\}_\varpi=\dd f(x)~.
\]
It can be checked
that the products defined in this way do form a Lie 2-algebra
structure on the complex $L:\,C^\infty(\Sigma)\rightarrow
C^\infty(M_1)$. One can also easily verify that these products satisfy
all the properties defining the exact twisted Courant algebroid $(T\Sigma\oplus
T^*\Sigma,\,\mu_2(-,-),\,\pi,\,H)$, where $\pi$ indicates the
algebroid's anchor map, that is the obvious projection to $\Sigma$,
and $H$ is the twisting 3-form. As we mentioned, we could include more
general twists (multi-vectors or mixed tensors), if needed. It is
further worth noting that this same discussion can be repeated for
higher Lie $n$-algebroids of the type $T^*[n]T[1]\Sigma$, carrying
higher Lie $n$-algebras, that will contain
the Vinogradov algebroid structures on spaces such as
$\Lambda^{n-1}T^*\Sigma\oplus T\Sigma$ \cite{Getzler:1010.5859, Zambon:2010ka}. These are of
interest when studying M-theory via exceptional generalized
geometry, when one wants to include 2- and 5-forms as fundamental
objects (see, for instance \cite{Hull:2007zu, Pacheco:2008ps}).

\subsection{Twisted Courant algebroids and $n$-plectic spaces}
We have now discussed two types of structures on potential space-time
manifolds $M$ that are somewhat new in theoretical physics, both born
out of a need to model certain properties of extended objects:
$n$-plectic spaces and exact Courant algebroids. Both were seen to be
examples of specific $n$-algebras. We will see now that they are in
fact related, since the latter contain the structure of the former, as
was shown in \cite{Rogers:2010sc}. 

Let us look again at the shlalo $\Pi_2$ of a $2$-plectic manifold
$M$:
\begin{equation*}
  \begin{array}{l}
    C^\infty(M)\stackrel{\pi_1}{\longrightarrow}\Omega^1(M)
    \qquad\text{with}\qquad \pi_1=\dd~,\\[1.5ex]
\pi_2(\alpha,\beta)=-\iota_{X_\alpha}\iota_{ X_\beta}\omega=\tfrac12(\iota_{X_\alpha}\dd\beta-\iota_{X_\beta}\dd\alpha)~,\\[1.2ex]
\pi_2(\alpha, f)=0~, \qquad
  \pi_3(\alpha,\beta,\gamma)=\iota_{X_\alpha}\iota_{X_\beta}\iota_{X_\gamma}\omega~.
  \end{array}
\end{equation*}
As with Lie algebras, we can define structure-preserving $n$-morphisms for
$n$-algebras (again, see \cite{Rogers:2010sc} for the $n=2$ case, and
\cite{Ritter:2015ffa} for the general discussion): in particular, there is a Lie 2-algebra isomorphic to
the shlalo described above, whose products are modified to the
following:
\begin{equation*}
  \begin{array}{l}
\pi_2(\alpha,\beta)=\tfrac12(\iota_{X_\alpha}\dd\beta-\iota_{X_\beta}\dd\alpha)+\dd\left(\iota_{X_\alpha}\beta-\iota_{X_\beta}\alpha\right)~,\\[1.2ex]
\pi_2(\alpha, f)=\iota_{X_\alpha}\dd f~, \qquad
  \pi_3(\alpha,\beta,\gamma)=\iota_{X_\alpha}\iota_{X_\beta}\iota_{X_\gamma}\omega~.
  \end{array}
\end{equation*}
Going back to the Lie 2-algebra corresponding to the Courant
algebroid: consider those degree-1 functions $e=X_\alpha+\alpha\in
C^\infty(M_1)$ whose vector field $X_\alpha$ is precisely the Hamiltonian
vector field corresponding to the 1-form $\alpha$, via the 2-plectic
structure $\omega=H$, the twisting 3-form of the Courant
bracket. Under this restriction, the Courant 2-algebra yields
precisely the above modified $\Pi_2$. Since the 2-plectic structure
$H$ needs to be non-degenerate, for the shlalo to make sense, we
cannot 'morph away' the twist 3-form, but it now allows for a more geometric
interpretation. 

It is also worth noting that all of the above
discussion can be generalized to higher $n$, to $\Pi_n$ on higher
dimensional spaces and its relation to the Vinogradov algebroids on
$T^*[n]T[1]M$, as is shown in \cite{Ritter:2015ffa}. Tying up all these
geometric and algebraic structures under the one common theme of
categorified algebras may lead to a deeper understanding, and/or
easier manipulation, in the context of double field theory and
possibly exceptional generalized geometry in M-theory.

\section{Example: effective M5-brane dynamics}\label{M5-brane section}
So far we have introduced the mathematical tools to describe gauge
theory based on Lie $n$-algebra internal symmetries, and we have seen
how some higher structures appear in geometry. We will now combine the
two: we construct and examine the equations of motion (\textit{i.e.}
the higher Maurer-Cartan equations) of a Lie 2-algebra model, but
rather than just on $T[1]\Sigma$, we will have it living on the
generalized space-time bundle $T^*[2]T[1]\Sigma$. That is, our
world-volume itself is now a graded $NQ$-manifold, with its own
higher $Q$-structure (as opposed to just $Q_\Sigma=\dd_\Sigma$). Interestingly, this approach contains the
same fields and yields the same dynamics as the proposal in
\cite{Lambert:2010wm} for the effective action of M5-branes. What
follows is a review of the detailed analysis presented in \cite{Ritter:2015zur}.
\subsection{Higher gauge theory on $T\Sigma\oplus T^*\Sigma$}\label{generalized
  higher gth}
To deduce the higher fake curvatures, we use the procedure
we elucidated in section \ref{Construction}. We are now looking at the diagram
\begin{equation*}
  \xymatrixcolsep{5pc}
  \myxymatrix{
  & T[1]L[1] \ar@{->}[d]^{\pi} \\
  T^*[2]T[1]\Sigma \ar@{->}[ur]^{f} \ar@{->}[r]^{a} & L[1]
  }
\end{equation*}
where we have the usual Lie 2-algebra
$L[1]:\,W[1]\stackrel{\,\mu_1}{\longleftarrow}V[2]$, with its $Q_L$-structure:
\[
Q_L=\left(-\frac12 f^a_{bc}w^bw^c
      -t^a_iv^i\right)\frac{\partial}{\partial
      w^a}+\left(\frac16 h_{abc}^iw^aw^bw^c-g_{aj}^iw^av^j\right)\frac{\partial}{\partial v^i}~.
\]
On the space-time side we use the
untwisted $Q_C$-structure 
\[
Q_C=\xi^\mu\der{x^\mu}+p_\mu\der{\xi_\mu}~,
\]
which, we recall from the previous section, gives the (untwisted) Courant algebroid
structure to $T[1]\Sigma\oplus T^*[1]\Sigma$. We now have coordinates of
degrees 1 and 2 respectively, $(w^a,\,v^i)$, on the $L[1]$ side, and
of degrees 0, 1 and 2, $(x^\mu,\,\xi^\mu,\,\xi_\mu,\,p_\mu)$, on the
$\Sigma$ side. That is, when using the grade preserving pullback
$a^*$, the most general 2-connection we obtain will be given by 
\begin{equation}
  \label{eq:1}
  \begin{aligned}
     A^a=a^*(w^a)&=A_\mu\xi^\mu+A^\mu\xi_\mu~,\\
B^i=a^*(v^i)&=\tfrac12 B_{MN}\xi^M\xi^N+B^\mu p_\mu~,
  \end{aligned}
\end{equation}
where the capital indices $M,\,N$ indicate both up and down $\mu$
indices, for compactness. We immediately notice the vector field $B^\mu$
at degree 2: its natural appearance in this framework is important to
the understanding of the proposal in \cite{Lambert:2010wm}, where such a
field is necessary, for the consistent behaviour of the theory under
dimensional reduction, but it is ultimately added in by hand. 

Recall that the higher curvatures were defined by the failure of $a$
to be a $Q$-morphism, as in eq. (\ref{curvature def}). Applying this
here, one obtains:
\begin{equation}
\begin{aligned}
      \CF^a&=\left[ \partial_MA_N+\tfrac12\mu_2(A_M,A_N)+\tfrac12\mu_1(B_{MN})\right]\xi^M\xi^N+\left(A^\mu+\mu_1(B^\mu)\right)p_\mu\\[1.4ex]
     \CF^i&=\left[-\tfrac16\mu_3(A_M,A_N,A_K)+\tfrac12\mu_2(A_M,B_{NK})+\tfrac12\partial_MB_{NK}\right]\xi^M\xi^N\xi^K\\[1.4ex]
\ &+\left(\mu_2(A_\mu,B^\nu)+B_\mu^\nu+\partial_\mu
      B^\nu\right)\xi^\mu p_\nu+\left(\mu_2(A^\mu,B^\nu)+\tfrac12
      B^{\mu\nu}\right)\xi_\mu p_\nu~\label{generalized higher curvature},
\end{aligned}
\end{equation}
for the part of the curvature $\CF^a$ valued in $W[1]$ and $\CF^i$ in
$V[2]$. Again, the capital indices $M,\,N$ run over upper and lower
$\mu$ indices, that is over all of $T[1]\Sigma\oplus
T^*[1]\Sigma$. 

We saw that the obvious topological higher gauge theory action
requires the higher curvatures to vanish, as its equations of
motion. We further know that this requirement really underlies the
whole motivation for the framework, since it is the only way to guarantee
a well-defined holonomy for extended objects. Indeed, it can be shown
that the equations of motion of various known supersymmetric theories
can be expressed as the vanishing of appropriately identified higher
curvatures (for detailed examples see \cite{Ritter:2015ymv}). It is
therefore reasonable to take this as the fundamental guiding
principle, even in the absence of an explicitly written action. In the
following subsection we will see how the above described
\textit{generalized 2-gauge theory}, via the zero-curvature principle,
reproduces precisely the equations of motion proposed by Lambert and
Papageorgakis for the effective dynamics of M5-branes.

\subsection{Effective dynamics of M5-branes}\label{M5-brane model}
Let us start by quickly reviewing the model proposed in \cite{Lambert:2010wm}. The
field content consists of the 6-dimensional $(2,0)$-multiplet: 5
scalar fields $X^I$, antichiral fermions $\Psi$ and the self-dual exact 3-form
$h=dB\in\Omega^3(\mathbb{R}^{1,5})$, all valued in
$\mathbb{R}^4$. Upon reduction along a circle to 5
dimensions, the supersymmetry transformation of these
fields should reproduce those of 5-dimensional super-Yang-Mills
theory, which include a term of the type $[X^I,X^J]$ for
$\delta\Psi$. The Ansatz chosen by the authors is to introduce a new
vector field $C=C^\mu\partial_\mu$, also valued in $\mathbb{R}^4$, to
couple to a term quadratic in $X^I$ in the 6-dimensional $\delta\Psi$,
which adjusts for the mismatch in chirality of $\Psi$ and the
supersymmetry parameter. The model further contains a gauge potential
$A_\mu\dd x^\mu$, valued in $\mathfrak{so}(4)$.

The internal symmetries $\mathbb{R}^4$ and $\mathfrak{so}(4)$ arise
because the authors' Ansatz is based on a 3-Lie algebra
structure\footnote{The nomenclature can be confusing, so we insist on
  reminding the reader that BLG 3-Lie algebras are not Lie 3-algebras.}, by
which we mean the ternary brackets first introduced by Bagger, Lambert and
Gustavsson (BLG) to model stacks of M2-branes
(cf. \cite{Bagger:2006sk, Gustavsson:2007vu}). As mentioned
earlier, these triple structures are in fact a particular example of
Lie 2-algebras: they correspond to \textit{strict} 2-algebras, that is
those whose Jacobiator $\mu_3$ is identically vanishing. Symmetry
considerations, like the closure of the superalgebra, together with
the correct behaviour under dimensional reduction, lead to the
following strict Lie 2-algebra for this model: the vector space complex
$L:\,W[1]\leftarrow V[2]$ has $W=\mathfrak{so}(4)$ and $V=\mathbb{R}^4$,
\[
L:\, *\leftarrow\mathfrak{so}(4)[1]\leftarrow\mathbb{R}^4[2]~.
\]
We are dealing with a strict 2-algebra, so the product $\mu_2$ on
$\mathfrak{so}(4)$ is just the Lie bracket $[-,-]_{\mathfrak{so}(4)}$ of the algebra, while the
action of $y\in\mathfrak{so}(4)$ on elements in $\chi\in\mathbb{R}^4$,
that is $\mu_2(y,\chi)$, is the obvious $\mathfrak{so}(4)$ action on vectors.
Before we move to the homotopy product $\mu_1$, we note that there
exists a map $D:\, \mathbb{R}^4\wedge\mathbb{R}^4\rightarrow
\mathfrak{so}(4)$ defined via
\[
\lbr y,\,D(\chi_1,\chi_2)\rbr_{\mathfrak{so}(4)}:=(y\chi_1,\chi_2)=-(y\chi_2,\chi_1)~,
\]
where $\lbr -,-\rbr$ stands for the metric on $\mathfrak{so}(4)$,
$(-,-)$ for that on $\mathbb{R}^4$ and $\chi_i\in\mathbb{R}^4$ while
$y\in\mathfrak{so}(4)$. The map $D$ can be used to construct the
antisymmetric BLG
ternary product: $[\chi_1,\chi_2,\chi_3]_{\text{BLG}}:=D(\chi_1,\chi_2)\chi_3$. It
is easy to check that this structure indeed satisfies the
\textit{fundamental identity} of BLG 3-Lie algebras (see also
\cite{Palmer:2012ya} for more details on the relation between BLG 3-algebras
and strict 2-algebras).

The conditions for the closure of the superalgebra of the M5-brane
model lead to the following equations of motion for the gauge gauge
fields: 
\begin{eqnarray}
0&=&h_{\mu\nu\kappa}-\tfrac{1}{3!}\varepsilon_{\mu\nu\kappa\rho\sigma\tau}h^{\rho\sigma\tau}\label{self-dual
  h}~,\\[1.3ex]
0&=& F_{\mu\nu}-D(C^\lambda,h_{\mu\nu\lambda})~,\label{curvature F}\\
0&=&\nabla_\mu C^\nu=D(C^\mu,C^\nu)~,\label{C squares to 0}\\
0&=& D(C^\rho,\nabla_r h_{\mu\nu\lambda})~,\label{C on h}
\end{eqnarray}
where the covariant derivative is given by $\nabla=\dd+A$ and
$F=\dd A+\tfrac12[A,A]_{\mathfrak{so}(4)}$ is the traditional curvature
of $A$. In the first line we just wrote the self-duality condition of
$h$ explicitly: as it turns out, $h$ could only be written as $\dd B$, for a 2-form
$B_{\mu\nu}$, if $B$ lived in a traditional abelian Lie algebra. Here,
however, though $B$ is valued in $\mathbb{R}^4$, it is part of the
more intricate 2-algebra structure and is always acted on by
$\mathfrak{so}(4)$-valued operators, thus carrying the non-abelian
structure with it. As a consequence, closure of the superalgebra over-constrains the
field and $h$ itself can no longer be exact, or interpreted as the
curvature of some 2-form (again, see \cite{Lambert:2010wm} for details). 

Before returning to this point,
however, we take a look at (\ref{C squares to 0}): the fact that
$D(C,C)=0$, implies that $C^\mu$ factorizes into a $c^\mu$ vector on
$\mathbb{R}^{1,5}$ and a constant $v\in\mathbb{R}^4[2]$. This means
that $D(v,-)$ now only spans an $\mathfrak{so}(3)$ subalgebra of
$\mathfrak{so}(4)$, which in turn implies that
$A\in\mathfrak{so}(3)$. Moreover, the map $D(v,-)$ can now be
interpreted as the homotopy map $\mu_1$, as it takes elements from
$\mathbb{R}^4[2]$ into elements in $\mathfrak{so}(3)[1]$ and is nilpotent. The strict Lie 2-algebra of interest therefore
reduces to
\[
L:\,*\stackrel{D(v,-)}{\longleftarrow}\mathfrak{so}(3)[1]\stackrel{D(v,-)}{\longleftarrow}\mathbb{R}^4[2]~.
\]

If we return our attention to the non-exact $h$, it turns out that
while it is not the traditional curvature of a 2-form field, it
can be re-expressed as a higher curvature. Indeed, introducing a
2-form $B$, such that $c^\mu B_{\mu\nu}=0$, defined via
\[
h_{\mu\nu\kappa}=\frac{1}{|c|^2}\left(B_{[\mu\nu}c_{\kappa]}+\frac{1}{3!}\varepsilon_{\mu\nu\kappa\lambda\rho\sigma}B^{[\lambda
    \rho}c^{\sigma]}\right)~,
\]
we can write its (strict) 2-curvature: 
\[
H=\dd B+\mu_2(A,B)=*H~,
\]
which can be checked to be self-dual (cf. \cite{Palmer:2012ya} for
details on this part). 

We would like the higher flatness conditions from our generalized
2-gauge theory, as described in section \ref{generalized
  higher gth}, to reproduce the equations of motion for the gauge
fields here. We know the 2-algebra structure we need for the internal
symmetry, while for the space-time side we set
$\Sigma=\mathbb{R}^{1,5}$, meaning that the $NQ$-manifold with the
appropriate Courant structure is $T^*[2]T[1]\mathbb{R}^{1,5}$.  We can
also identify our vector field $B^\mu$ arising as a pullback of a
degree 2 object in $L$ with the field $C^\mu$ appearing in the
M5-brane model. This means that we can impose that $D(B^\mu,B^\nu)=0$. What
happens if we set the the higher curvatures (\ref{generalized higher
  curvature}) to zero? Let us start by those components that are
proportional to $p_\mu$. Since $\mu_1(B^\mu)\sim D(B^\nu,B^\mu)$, we
deduce from the $p_\mu$ term in $\CF^a$ that $A^\mu=0$. From the last
term in $\CF^i$, this implies that also $B^{\mu\nu}=0$. If we now look
at $(\CF^i)_{\mu\nu}^{\phantom{\mu\nu}\rho}$, we see that its
vanishing requires $B_\nu^{\phantom\nu\rho}$ to be covariantly
constant, so that it can be gauged away. Now, the only non-zero components left in
$(\CF^a)_{MN}$ are $\CF^a_{\mu\nu}$, whose vanishing gives
(\ref{curvature F}). Requiring $(\CF^i)_\mu^{\phantom\mu \nu}=0$ is
just the first part in equation (\ref{C squares to 0}), stating that
$\nabla_\mu C^\nu=0$. The last condition that is left is just the
vanishing of the component $(\CF^i)_{\mu\nu\rho}$, which is precisely
the self-duality condition for $h$. We have therefore recovered all
the equations of motion of the gauge fields of the M5-brane model,
using just our guiding principle of setting higher curvatures to zero,
albeit, here, with a generalized space-time bundle with its own Lie
2-algebra structure. The novel vector field $C^\mu=B^\mu$ finds a
natural role in the higher gauge-theory framework.

\section{Conclusions}

Recapping, we argue that the fundamental symmetry structure for any
type of gauge theory of extended objects is given by truncated
L$_\infty$-algebras, or Lie $n$-algebras. We have shown how to
construct topological higher gauge theories, and we have seen how
such models are solved by gauge $n$-connections that satisfy higher
flatness conditions. These conditions are also the ones that guarantee
that one can well define a concept of holonomy, or of ``surface
ordering'', over extended world-volumes. Naturally, this is a
fundamental requirement if we want to write down reparametrization
invariant theories. We therefore consider this same set of conditions
to be our guiding principle for describing any kind of model involving
extended fundamental objects, not only topological ones. Indeed, the
equations of motion of various supersymmetric theories can be
rewritten as higher flatness conditions, as can be seen in
\cite{Ritter:2015ymv}. In particular, here we have shown the details of how this works
for the M5-brane model proposed by Lambert and Papageorgakis \cite{Lambert:2010wm}.

This model for M5-branes is actually equivalent to one
of D4-branes, as for any choice of vacuum expectation value for
$C^\mu$, the total symmetry group breaks down to that expected for
5-dimensional super-Yang-Mills theory. We are not arguing that this discussion be complete, but we find it
interesting and possibly quite meaningful that even the simplest
higher structure can give insight not only into more complex behaviour
of gauge fields, but also into the field content itself necessary for
a consistent theory. It is of course seductive to think that the
generalized geometry introduced to better understand the dualities
of an extended object, such as a string, is just another
manifestation of the higher gauge structures underlying a more
complete quantum theory. 

We showed how, along with Courant algebroids, also the BLG triple
brackets, as well as the natural structure on $n$-plectic manifolds,
all fit into the same higher algebra framework. When looking to
quantize extended space-times themselves, we therefore propose that the latter
$\Pi_n$ algebras are the correct objects to be considering, where
Nambu-Poisson manifolds could be too restrictive. It seems that the best way to
approach this issue is via some higher geometric quantization, as is
being investigated by \cite{Rogers:2010sc, 1106.4068, 1304.6292,Fiorenza:2013kqa}.

Overall, the unifying power of the higher gauge algebra framework
should definitely be a step forward in the quest for a theory of
quantum gravity. Indeed, strong homotopy algebras appear in certain
reformulations of gravity \cite{Baez:2012bn}, and they are expected to underly spin-foam
models \cite{Baez:1997zt, Crane:2003jq, Baratin:2009za}. Furthermore, they made one of their first
appearances in physics in the context of string field theory
\cite{Zwiebach:1992ie} and it may not be surprising that they may encode the full symmetry structure of higher spin
theories as well \cite{Berends:1984rq, Stasheff:1993ny, Bengtsson:2008mw}. We would therefore like to leave
the reader with the thought that, possibly, unifying all these
approaches into the one appropriate language or formalism,
might bring the construction of our scientific tower of Babel one
step closer. 

\bibliographystyle{ws-procs9x6} 

\end{document}